\documentclass{moriond}

\newcommand\farcs{\hbox{$.\!\!^{\prime\prime}$}}
\def\ga{\mathrel{\mathchoice {\vcenter{\offinterlineskip\halign{\hfil
$\displaystyle##$\hfil\cr>\cr\sim\cr}}}
{\vcenter{\offinterlineskip\halign{\hfil$\textstyle##$\hfil\cr
>\cr\sim\cr}}}
{\vcenter{\offinterlineskip\halign{\hfil$\scriptstyle##$\hfil\cr
>\cr\sim\cr}}}
{\vcenter{\offinterlineskip\halign{\hfil$\scriptscriptstyle##$\hfil\cr
>\cr\sim\cr}}}}}

\usepackage{natbib}
\usepackage{url}

\newcommand{\Photo}{}

\begin{document}
\vspace*{4cm} \title{An overview of the completed Canada-France-Hawaii
  Telescope Lensing Survey (CFHTLenS)}

\author{ H. Hildebrandt on behalf of the CFHTLenS collaboration }

\address{Argelander-Institut f\"ur Astronomie, Universit\"at Bonn, Auf dem H\"ugel 71, 53121 Bonn, Germany}

\maketitle\abstracts{The Canada-France-Hawaii Telescope Legacy Survey
  (CFHTLS) represents the most powerful weak lensing survey carried
  out to date. The CFHTLenS (Canada-France-Hawaii Telescope Lensing
  Survey) team was formed in 2008 to analyse the data from the CFHTLS
  focussing on a rigorous treatment of systematic effects in shape
  measurements and photometric redshifts. Here we review the technical
  challenges that we faced in analysing these data and their solutions
  which set the current standard for weak lensing analyses. We also
  present some science highlights that were made possible by this
  effort including cosmic shear tomography, tests for modified gravity
  models, and the mapping of dark matter structures over
  unprecedentedly large scales. An outlook is given on current and
  future surveys that are analysed with the tools prepared for
  CFHTLenS. {\bf CFHTLenS represents the first and only weak lensing
    data set that has been made publicly available so far.} We
  encourage other surveys to follow this example.}

\section{Introduction}
The first observations of cosmic shear
\citep{2000MNRAS.318..625B,2000A&A...358...30V,2000Natur.405..143W}
opened up a new window to study cosmology. Measuring the build up of
the large scale distribution of dark matter structures over cosmic
time with cosmic shear tomography represents a very powerful
cosmological tool that ideally complements the high-redshift
observations of the cosmic microwave background (CMB) with
observations at low-redshift. Large volumes have to be surveyed to
yield cosmologically meaningful results, and the three-dimensional
distribution of structures has to be resolved for the tightest
cosmological constraints. These requirements naturally lead to the
design of deep multi-band imaging surveys over large areas of the sky
with the data being taken under the best possible seeing
conditions. The multi-band data are important for estimating
photometric redshifts of millions of background galaxies that are used
for the lensing measurement whereas the high-resolution is crucial for
measuring accurate ellipticities for the same objects.

The Canada-France-Hawaii Telescope Legacy Survey (CFHTLS) is the most
ambitious project with such a focus that has been completed so
far. Here we describe the work of the CFHTLenS (Canada-France-Hawaii
Telescope Lensing Survey) team that was formed to analyse the data
from the CFHTLS focussing on a rigorous treatment of the main sources
of systematic errors in such an analysis. In Sect.~\ref{sec:cfhtls} we
describe the CHFTLS data set and in Sect.~\ref{sec:analysis} we
illustrate some of the technical challenges that we were faced with
analysing those data. Section~\ref{sec:science} presents a few
highlights from the scientific papers that were made possible by the
unprecedented quality of the CFHTLenS data products. In
Sect.~\ref{sec:summary} we give an outlook and report on some projects
that are currently carried out with the tools developed for CFHTLenS.

\section{The CFHTLS}
\label{sec:cfhtls}
CFHTLenS is based on the Wide part of the CFHTLS, an imaging survey
carried out with the MegaCam instrument mounted at the CFHT. It
consists of 171 pointings of this 1 square degree camera, which are
arranged in four different contiguous patches to allow observations
all year. All patches are located at high-galactic latitude and three
of them are close to the equator whereas one is at higher
declination. All pointings have been observed in the $ugriz$-bands to
a 5$\sigma$ depth of 25.2, 25.6, 24.9, 24.6, and 23.5,
respectively. The best seeing time is reserved for the $i$-band data
so that the data from this band with a median seeing of $0\farcs7$ are
ideally suited for weak lensing shape measurements.

\section{Data analysis}
\label{sec:analysis}
\subsection{The CFHTLenS team}
The CFHTLenS team was formed to re-analyse the CFHTLS data with the
goal of understanding all systematic effects that affect cosmic shear
science to a level that is better than the statistical precision of
this data set. To reach this goal we apply and further develop the
most advanced shape measurement and photometric redshift (photo-$z$)
algorithms. The team originally consisted of members from different
European countries and Canada which had been working together on the
STEP shape measurement challenge \citep{2006MNRAS.368.1323H} and other
projects.

\subsection{Data reduction}
All MegaCam data are pre-reduced at the Canadian Astronomical Data
Centre (CADC). We are using these pre-reduced individual exposures as
the basis for our data reduction. Subsequent reduction steps are
carried out with the THELI wide-field imaging reduction pipeline
\citep{2003A&A...407..869S,2005AN....326..432E}. These include
astrometric and photometric calibration, stacking, masking, and
creation of weight and flag images. Details can be found in
\cite{2013MNRAS.433.2545E}. While the deep stacks are used for the
extraction of multi-colour photometry we measure galaxy shapes from
individual exposures to avoid a number of systematic effects that are
created by the stacking procedure.

\subsection{Photometry and photo-$z$}
The point spread function (PSF) is typically different in the
different bands of a pointing. In order to extract accurate colours
for all objects this has to be accounted for. We convolve the five
images (i.e. the $ugriz$ stacks) of one pointing so that the PSF in
all bands is Gaussian and has the same size. This is done with the
shapelet-based code described in \cite{2008A&A...482.1053K}.

The photometry is then extracted from these Gaussianised stacks with
SExtractor \citep{1996A&AS..117..393B} in dual-image mode. The
unconvolved $i$-band image is used for detection. This procedure
yields unbiased colours with close to optimal signal to noise ratio
(S/N).

Photometric redshifts are then estimated with the BPZ photo-$z$ code
\citep{2000ApJ...536..571B} and the results are compared to different
spectroscopic redshift catalogues that are overlapping with the CFHTLS
footprint. Details of the multi-colour photometry and photo-$z$
methods can be found in \cite{2012MNRAS.421.2355H}.

\subsection{Shape measurement}
A particular focus within CFHTLenS is given to shape measurements
\citep{2013MNRAS.429.2858M} since this area had been identified as
being affected by different systematic effects. In particular redshift
dependent systematics \footnote{Those can originate from flux/size
  dependent systematics and the correlation between redshift and
  flux/size.} are harmful for cosmic shear tomography, one of the main
science drivers for CFHTLenS. While we started with a number of
different shape measurement techniques we quickly concentrated on the
Bayesian \emph{lens}fit algorithm
\citep{2007MNRAS.382..315M,2008MNRAS.390..149K} that showed the most
promise to meet our requirements.

\emph{Lens}Fit is a Bayesian forward fitting shape measurement code that
employs a suite of analytical galaxy light profiles convolved with the
measured PSF to fit the data. The resulting multi-dimensional
likelihoods are then multiplied with empirical prior distributions for
galaxy size, ellipticity, and bulge to disk ratio to yield a posterior
probability distribution. Marginalising over all uninteresting
parameters then yields an unbiased estimate of the ellipticity of each
galaxy. Furthermore, a weight is calculated that parametrises the
error of the ellipticity measurement and can readily be used in shear
measurements to properly weigh background sources.

As mentioned above \emph{lens}fit is run on the individual,
astrometrically-calibrated exposures. Results from different exposures
of one field (typically 7 per field) are combined in a statistically
optimal way. Working on individual exposures instead of stacks has the
advantage that different PSFs are not mixed. Besides that, it also
avoids the correlation of noise that is an inevitable result of
sub-pixel stacking.

\subsection{Tests for shape systematic}
An important aspect of the work of the CFHTLenS team is the
development of cosmology-independent systematic tests to check the
robustness of the shear catalogue. 

The most important tool to identify residual effects from an imperfect
PSF correction is the star-galaxy cross-correlation function
\citep{2012MNRAS.427..146H}. Here, the shapes of the corrected
galaxies are cross-correlated with the shapes of the uncorrected stars
that were used to measure the PSF. Naively one could assume that a
signal significantly different from zero would indicate the presence
of residual systematics. However, not only shot noise contributes
here, and the significance has to be estimated taking into account a
signal that can be created by cosmic shear itself \citep[see][for
details]{2012MNRAS.427..146H}. This means that the amplitude of the
star-galaxy cross-correlation function has to be compared to results
from detailed simulations with - by construction - perfect PSF
correction. We find that the amplitude of the star-galaxy
cross-correlation function for all fields is considerably higher than
the one found in the simulations. However, this undesired signal is
dominated by a few fields that seem to have larger systematic errors
than others. Rejecting $\sim25\%$ of the fields with the strongest
star-galaxy cross-correlation leads to a star-galaxy cross-correlation
function amplitude consistent with simulations.

We also run \emph{lens}fit on detailed image simulations that are carefully
matched to the data in terms of the distributions in ellipticity, S/N,
size, and bulge to disk ratio \citep{2013MNRAS.429.2858M}. This is
done to test the output ellipticities $e_1$ and $e_2$ against the
known input ellipticities as functions of S/N and
size (both of which can be relatively easily be measured from the data
itself). This reveals some residual multiplicative biases that we
correct for on a object-by-object basis in the data. Note that this
kind of bias is expected in noisy measurements and can not be
completely avoided.

The image simulations do not reveal any additive bias. However,
averaging the $e_2$ ellipticity component for all CFHTLenS data
reveals some residual bias that is significantly different from
zero. We characterise this bias as a function of size and S/N and
subtract it from the measured $e_2$ \citep{2012MNRAS.427..146H}.

The combination of rejecting fields with bad PSF (identified by their
anomalously high star-galaxy cross-correlation) and the correction of
all residual biases with image simulations and directly from the data
yields a state of the art shear catalogue that is free from systematic
errors to the level required by the size of the CFHTLS.

\section{Science highlights}
\label{sec:science}

\subsection{Cosmic shear}
One of the main science drivers of CFHTLenS is cosmic shear
tomography. Measuring the statistical properties of the large scale
structure of the dark matter density field over cosmic time has become
one of the most promising cosmological tools. This cosmological probe
is studied in great detail in a number of CFHTLenS publications
\citep{2013MNRAS.430.2200K,2013MNRAS.431.1547B,2013MNRAS.432.2433H,2014arXiv1401.6842K,2014arXiv1404.5469F}.

A pure 2D cosmic shear analysis is presented in
\cite{2013MNRAS.430.2200K} where the individual redshifts of the
sources are not taken into account and the colour information are
purely used to constrain the redshift distribution. In this paper we
explore different ways of measuring the signal from small, non-linear
scales out to large radii ($\ga2^\circ$). It is shown how such
measurements yield tight constraints on the total cosmic matter
density, $\Omega_{\rm m}$, the amplitude of the fluctuations of the
matter power spectrum, $\sigma_8$, and the curvature, $\Omega_{\rm
  K}$. These constraints are complementary to constraints from the CMB
and improve on what has been found in the seven-year data set of the
WMAP satellite \citep{2011ApJS..192...18K}. An extension of this 2D
cosmic shear measurement to third-order statistics can be found in
\cite{2014arXiv1404.5469F}. It is shown that these higher-order
statistics add some statistical power to the cosmic shear result at
the expense of more complicated systematic errors.

For tomographic cosmic shear we split up the source sample into
different redshift bins. A basic tomography study with two broad
redshift bins and concentrating on the systematic robustness of the
photo-$z$ is presented in \cite{2013MNRAS.431.1547B}. A much more
detailed cosmic shear tomography measurement is presented in
\cite{2013MNRAS.432.2433H}. There we split up the source sample into
six narrower redshift bins and measure all 21 possible shear cross-
and auto-correlation functions. In this study we also correct for the
major astrophysical systematic associated with cosmic shear
measurements: intrinsic alignments of galaxy ellipticities, which are
especially important at low redshift. The inclusion of redshift
information yields tighter constraints on the cosmological parameters
discussed above and allows for testing the dark energy equation of
state, $w$. Assuming flatness and combining our results with different
external data sets we constrain this crucially important cosmological
parameter to $w=-1.02\pm0.09$.

Instead of binning galaxies in discrete photo-$z$ bins one can also
use the redshift information for each galaxy individually. This is
known as 3D cosmic shear and an application to CFHTLenS is presented
in \cite{2014arXiv1401.6842K}. Taking into account the full redshift
probability distribution of each object that is provided by the
photo-$z$ code we can suppress non-linear scales much more
rigorously. This is advantageous because it means that we can compare
our measurements to theory in a regime where cosmic structure
formation is better understood than on the small scales where baryonic
effects and non-linear evolution play a great role. Hence, the
uncertainty of the models is considerably reduced. It is clear that
precision suffers if the small scales with high S/N are
neglected. However, this is not a problem of 3D cosmic shear itself
(it could easily be used on smaller scales) but rather stresses the
importance of better modelling these small-scale effects for future
dark energy mission which will have considerably greater statistical
power.

\subsection{Modified gravity}
One of the most interesting aspects of cosmic shear tomography is that
it is sensitive to both measurable effects of dark energy, the
influence dark energy has on the geometry of the Universe as well as
on the growth of the large scale structure. It is the combination of
these two different aspects that has the potential to reveal
deviations from general relativity and test alternative theories of
gravity. This becomes even more powerful when cosmic shear is combined
with a tracer of non-relativistic physics. In
\cite{2013MNRAS.429.2249S} we use redshift-space distortions (RSD)
from the WiggleZ dark energy survey \citep{2010MNRAS.401.1429D} and
6dFGS \citep{2009MNRAS.399..683J} for that purpose. Combining these
RSD measurements with the 2-bin cosmic shear tomography results from
\cite{2013MNRAS.431.1547B} we obtain constraints on several
parametrisations of modified gravity models. All our constraints are
compatible with general relativity excluding large parts of the
parameter space for possible deviations. This study illustrates the
power of this combination to potentially falsify one of the pillars of
our physical world model, and such measurements are the basis for very
large projects that are starting just now (e.g. 2dFLens).

\subsection{Dark matter maps}
The large contiguous patches of CFHTLenS and the high-quality data
resulting in high number densities of lensing sources represent ideal
conditions for mapping the dark matter distribution directly. This
creation of mass maps and their scientific exploitation is described
in \cite{2013MNRAS.433.3373V}. The value of these maps goes beyond the
mere visualisation of dark matter structures. Maps are especially
useful to study higher-order statistics and the relationship between
luminous and dark matter \citep[see also][]{2013MNRAS.433.3373V}.  The
CFHTLenS dark matter maps are the largest created so far and reveal
giant voids on scales of several degrees that were undetectable with
previous surveys. Furthermore, the maps offer the exciting opportunity
to cross-correlate the dark matter structures to signals extracted
from very different experiments \citep[for an example
see][]{2014PhRvD..89b3508V}.

\section{Summary and outlook}
\label{sec:summary}
The work of the CFHTLenS team represents the next crucial step in
controlling systematic effects in state of the art weak gravitational
lensing surveys. The analysis tools developed in the course of the
project go a long way beyond what has been used in the past and form
the basis for current and upcoming Stage III and IV weak lensing
projects that are 1-2 orders of magnitude larger than CFHTLS and
require much tighter control of systematic errors.

In certain areas (dark energy equation of state, modified gravity
parameters, dark matter mapping) the data analyses from CFHTLenS yield
competitive or even unprecedented results showing the full potential
of weak lensing as a cosmological tool. The scientific relevance of
CFHTLenS goes beyond pure cosmological results, and the data have been
used by the CFHTLenS team for studies of galaxy evolution
\citep{2013MNRAS.430.2476S,2013arXiv1310.6784H,2014MNRAS.437.2111V} as
well as galaxy group and cluster science
\citep{2010MNRAS.406..673M,2013MNRAS.431.1439G,2014MNRAS.439.3755F}.

The CFHTLenS data are publicly available at
\url{http://www.cfhtlens.org} as well as at
\url{http://www.cadc-ccda.hia-iha.nrc-cnrc.gc.ca/community/CFHTLens/query.html}

We provide calibrated images, photometry and shear catalogues, masks,
random catalogues matching the geometry of the data, cosmological data
vectors for several of the measurements discussed here, covariance
matrices, and redshift distributions. The data have been downloaded by
a large number of different researchers (as evidenced by the number of
unique IP addresses), and a growing number of external papers using
CFHTLenS data are coming out. {\bf It should be noted that CFHTLenS is
  the only weak lensing data set that has been released publicly so
  far.} We strongly encourage the community to follow our example and
support the credibility of weak lensing as a cosmological tool in
general and our understanding and treatment of systematic effects in
particular through open access of weak lensing data.

The suite of tools developed within CFHTLenS is currently being
applied to a number of similar surveys. CFHTLenS members are strongly
involved in the European Kilo Degree Survey
\citep[KiDS][]{2013ExA....35...25D} as well as in a re-analysis of the
data from the Red Sequence Cluster Survey 2 data (this project is
dubbed RCSLens\footnote{\url{http://www.rcslens.org}}). Combined with
similar data analysis on smaller projects like CS82 (CFHT Stripe 82
Survey), NGVS \citep[The Next Generation Virgo
Survey]{2012ApJS..200....4F}, and CODEX (weak lensing calibration to
Constrain Dark Energy with X-ray clusters) this will yield a state of
the art lensing data set of $\sim 3000{\rm deg}^2$ in the near future
rivalling a project like the Dark Energy Survey in area and surpassing
it in image quality. We feel committed to make also these data
publicly available once the main science analyses by the teams are
done.

\bibliography{2014-03_Moriond_Hildebrandt}

\end{document}